\documentclass[]{zakoproc}
\usepackage{graphicx}
\usepackage{natbib}     % ADDED natbib package
\begin{document}

%\title{SMA High-Resolution Magnetic Field Observations and Analysis Methods}
\title{Observations and Analysis of High-Resolution Magnetic Field Structures in Molecular Clouds}
\author{Ya-Wen Tang$^1$, Patrick M. Koch$^2$ \& Paul T.P. Ho$^2$}
\institute{$^1$ Univ. Bordeaux LAB and CNRS, UMR 5804, F-33270, Floirac, France\\
$^2$ Academia Sinica, Institute of Astronomy and Astrophysics, Taipei, Taiwan}
\markboth{Y.-W. Tang et al.}{Magnetic Field Observations and Analysis}

\maketitle

\begin{abstract}
Recent high-angular-resolution (up to $0\farcs7$) dust polarization observations toward star forming regions are summarized. With the Sub-Millimeter Array, the emission from the dense structures is traced and resolved.
The detected magnetic field morphologies
vary from hourglass-like structures to isolated patches depending on the evolutionary
stage of the source. These observed features 
have also served as a testbed to develop new analysis methods, with a particular 
focus on quantifying the role of the magnetic field in the star formation process.
\end{abstract}

\section{Introduction}

The connection of the magnetic field (B) in star formation regions from the molecular cloud and envelope (typically at a scale of 1 pc) down to the collapsing core (at a scale of a few mpc) is not yet clear. 
%Polarization in the diffuse material at the scale of several pc have been detected in the optical. 
It has only been possible to resolve the B field in dense cores in the past decade with 
the recent advancing instrumental developments, namely the high-angular resolution, sensitivity and the capability of detecting the (sub)millimeter emission. 
The emission at (sub)millimeter is particularly interesting, because it is mostly optically thin and allows us to trace dense structures.
Here, we present a sample of our measurements where the sources are at different evolutionary stages, 
from the collapsing core (W51 e2 and part of Orion BN/KL) to the ultra-compact HII region G5.89-0.39.
The observations were carried out using the Sub-Millimeter Array (SMA) in various array configurations. 
Dust continuum and its polarized emission at 0.88 mm were imaged. With the SMA, dense structures with the number density 10$^{5}$ to 10$^{7}$ cm$^{-3}$ are traced.
The plane-of-sky projected B field 
integrated along the line of sight, B$_{\bot}$, is derived by rotating the detected polarization by 90$\degr$, 
assuming that the dust grains are aligned with their shorter axes parallel to the field lines. 

\section{High-Resolution Polarization Observations}

\begin{figure}[h!]
%../../RESEARCH/SMA_W51/SMA_W51_080713/WIP_MAPS_W51_0_TOTAL-self/sma_vec_e2_2and3sig2.eps
\includegraphics[scale=0.43]{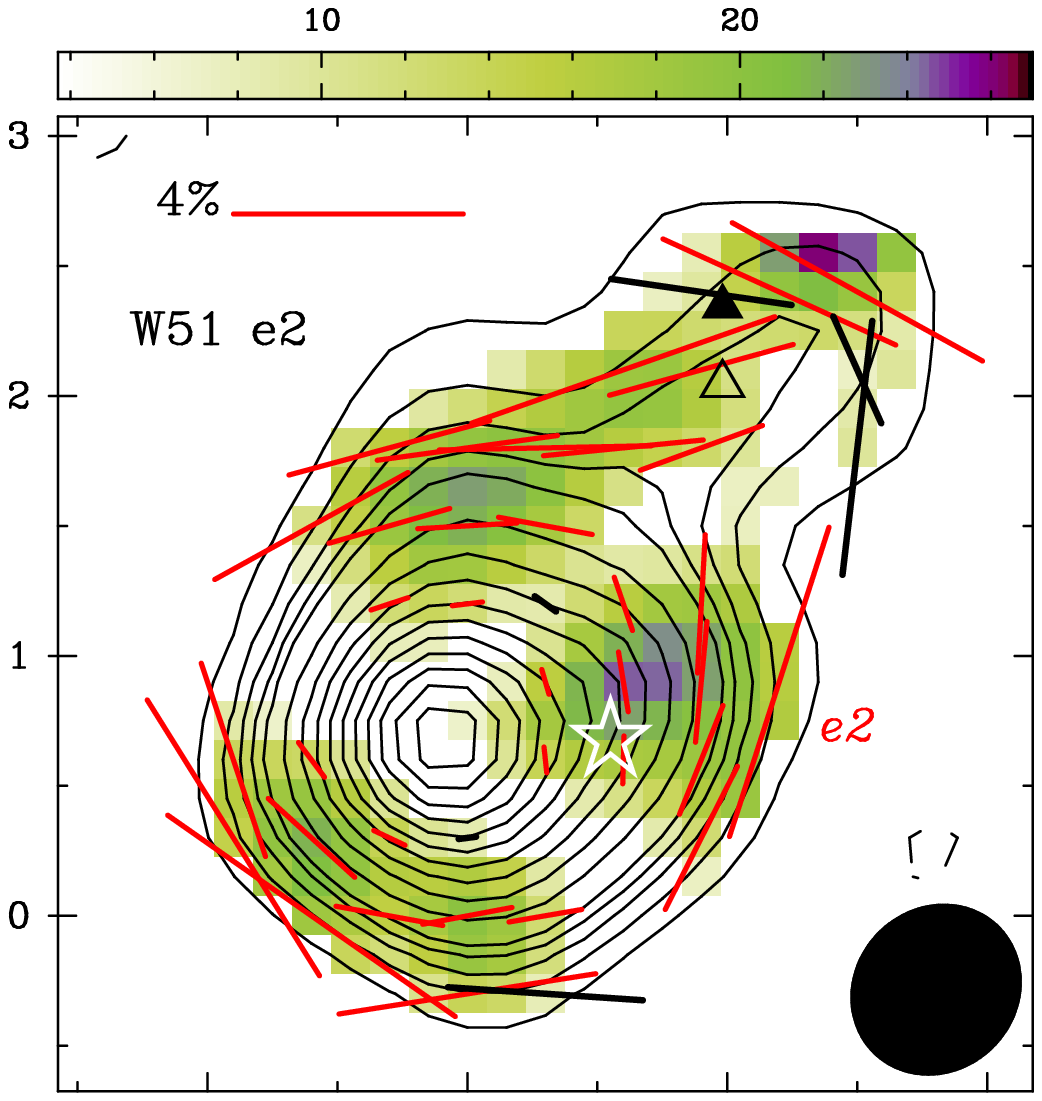}
%../../RESEARCH/POLMAPS/ip_dust_cont_pol2.eps
\includegraphics[scale=0.43]{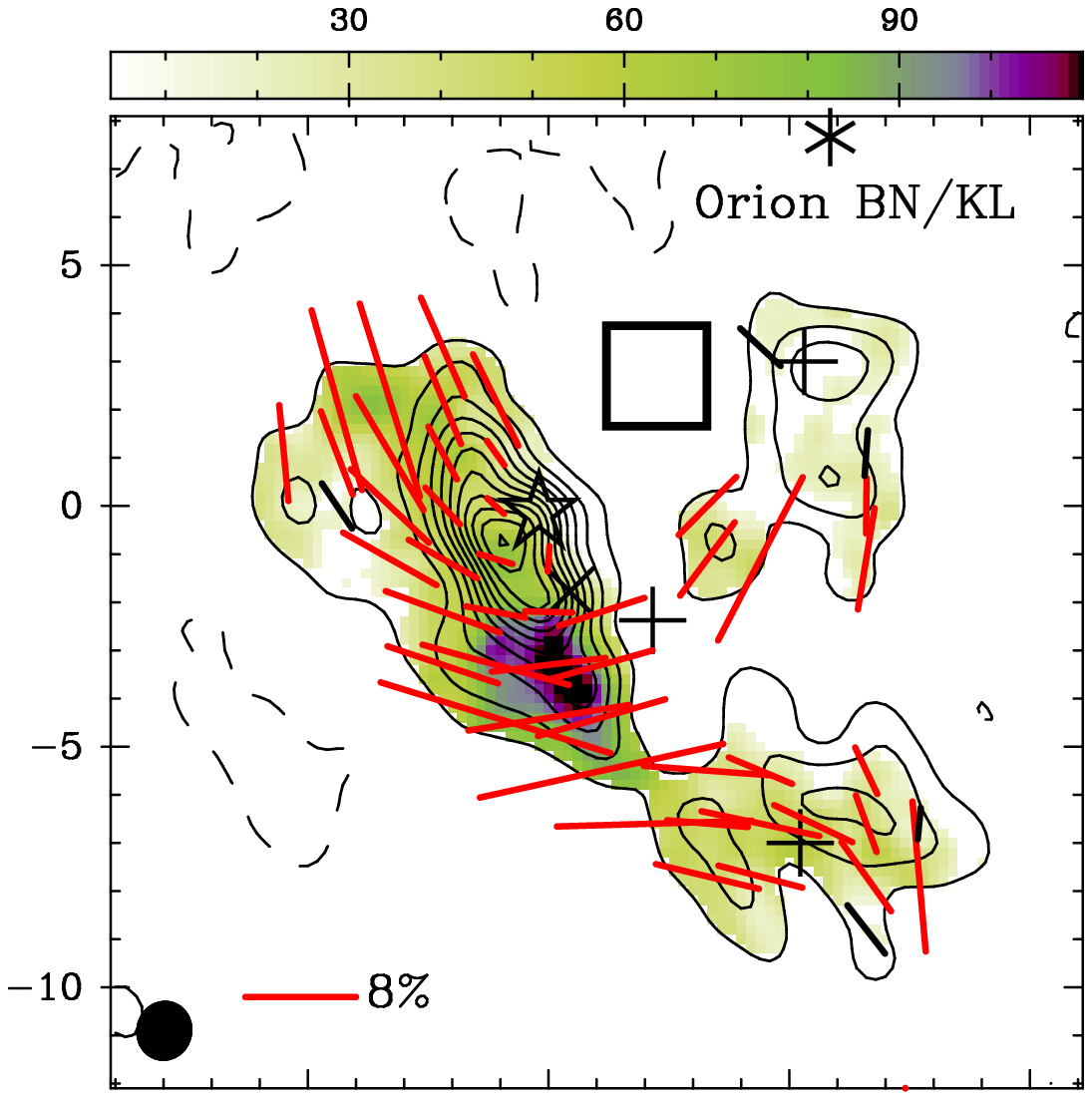}
%../../RESEARCH/G589/g589_combself/overlay/ip_dust_cont_pol2.eps
\includegraphics[scale=0.43]{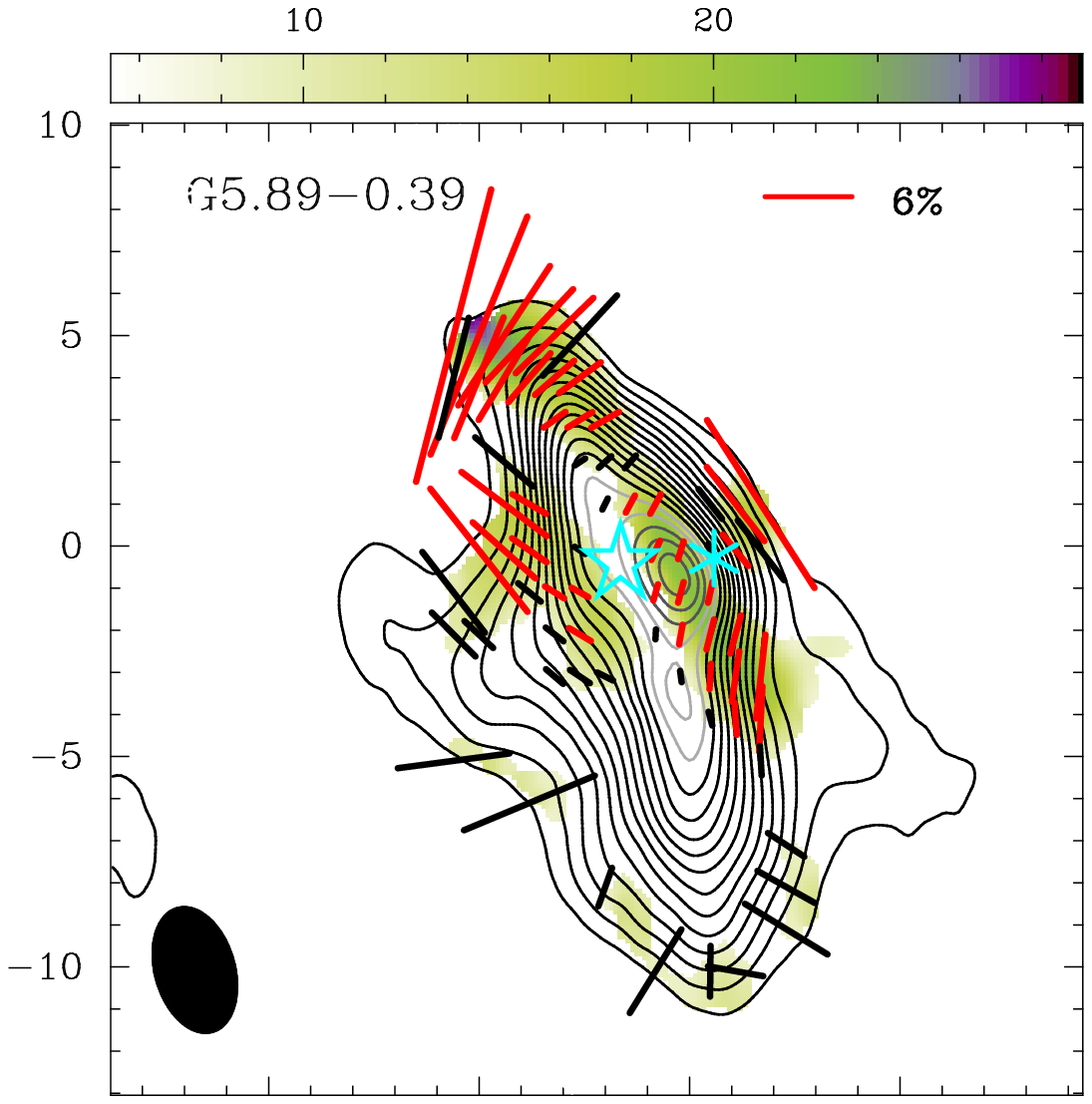}
%../../RESEARCH/SMA_W51/SMA_W51_080713/WIP_MAPS_W51_0_TOTAL-self/sma_B_e2_2and3sig2.eps
\includegraphics[scale=0.43]{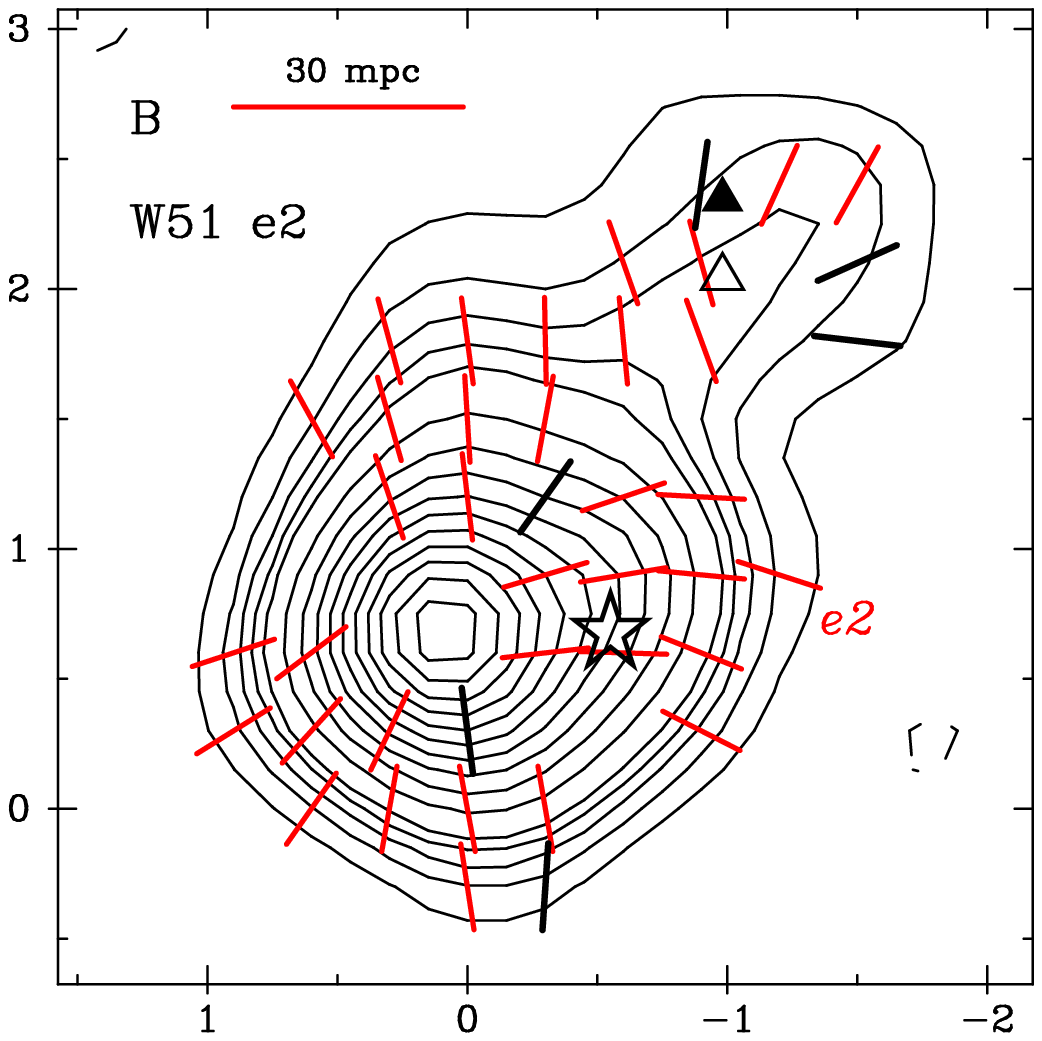}
%../../RESEARCH/POLMAPS/b_dust_cont_pol2.eps
\includegraphics[scale=0.43]{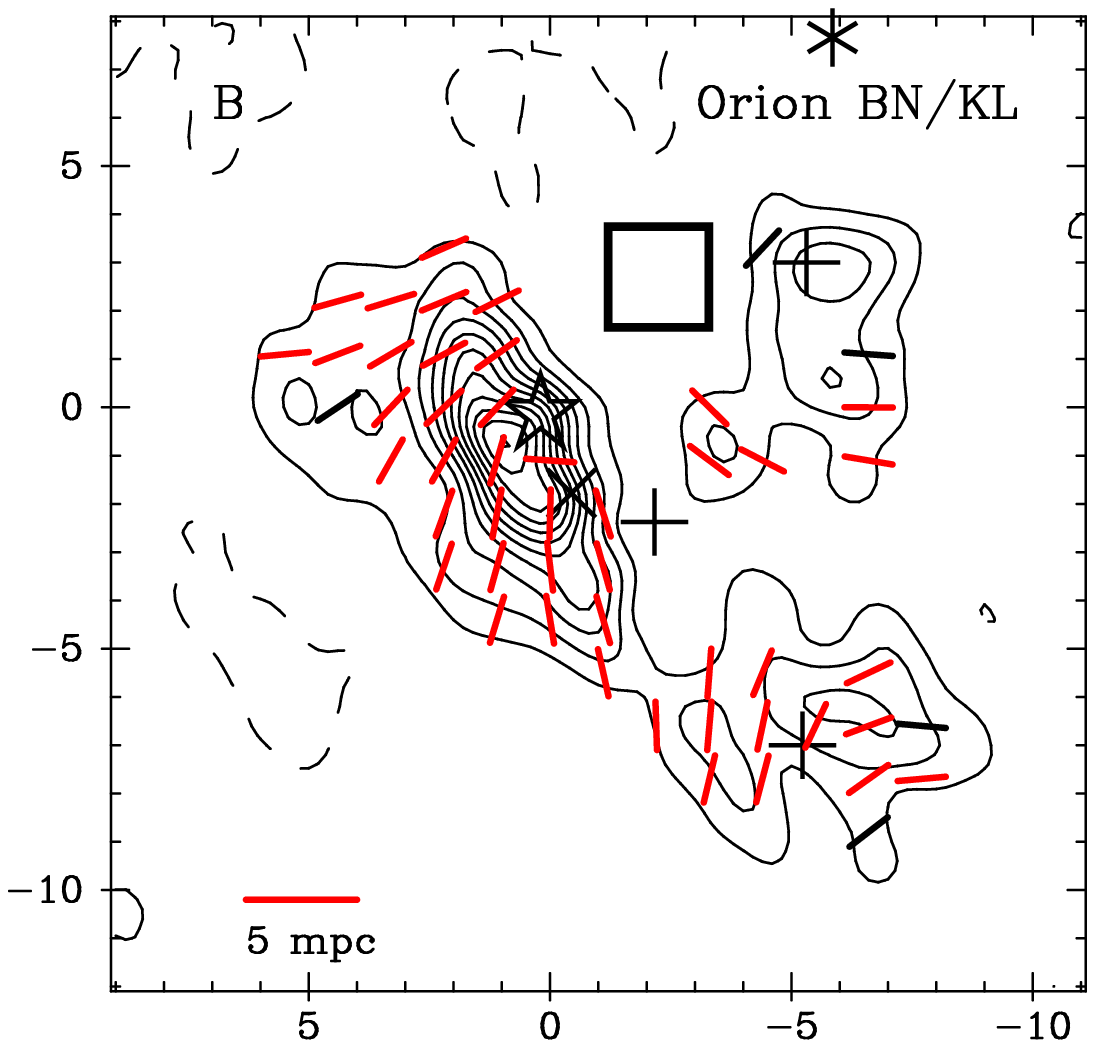}
%../../RESEARCH/G589/g589_combself/overlay/b_dust_cont.eps
\includegraphics[scale=0.43]{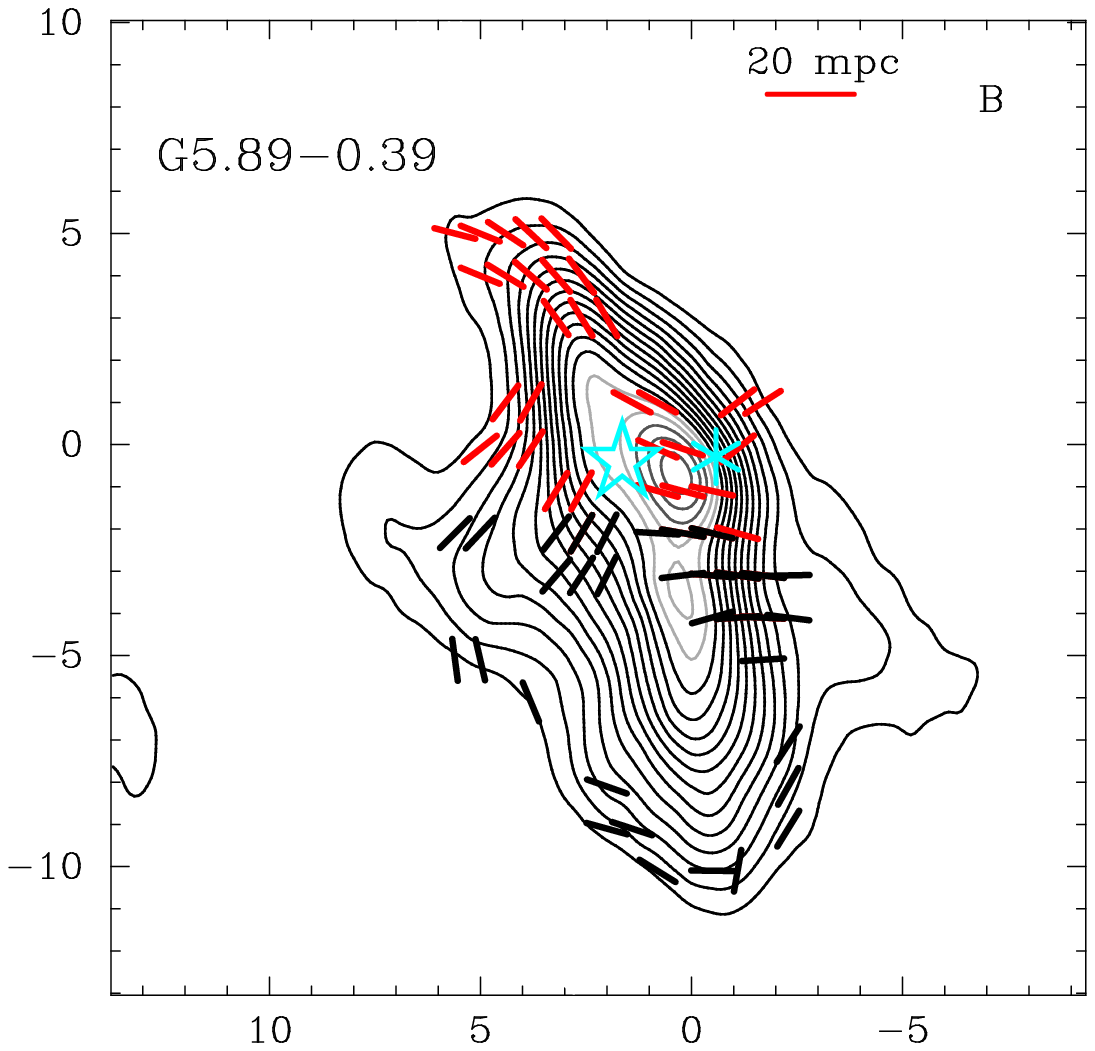}
\caption{\footnotesize Upper panels: Polarization intensity (color scale) in mJy beam$^{-1}$ and polarization 
above 3 $\sigma_{\rm Ip}$ and between 2 to 3 $\sigma_{\rm Ip}$ in red and black segments, respectively. 
The black contours represent the emission strength of the Stokes I component. Lower panels: B field map (segments) 
of the corresponding sources. The axes are the offsets from the original phase centers in Right Ascension and Declination in arcsecond. Note that the B segments are derived by rotating the detected polarization by 90$\degr$ with equal length. The black ellipses mark the angular resolution.}
\label{description}
\end{figure}

%\subsection{Description of individual sources}
Based on the kinematics derived from molecular lines, the collapsing core W51 e2 was found to be embedded in a 
gravitationally stable envelope. 
The B field in the envelope has been mapped by Lai et al. (2001) 
revealing a uniform field morphology. 
With an angular resolution, $\theta\approx0\farcs$7, the field within the core was resolved with 
the SMA (left panels in Figure 1 and \citet{tang09b}). The coincidence of de-polarization zones with 
the ionized accretion disk and the molecular outflow suggests that the B field is associated with 
the accreting material. An hourglass-like B$_{\bot}$ morphology is found. 

The B field in Orion BN/KL appears to be more complex. The polarization varies significantly but 
smoothly across the detected patches (middle panels in Figure 1 and \citet{tang10}). Two symmetry features have been found: 
firstly, the B field shows a radial symmetry, where the center is within the uncertainty of the origin of 
the explosive molecular outflows; secondly, there is a symmetry plane, 
with mirror symmetric field features on both sides. This plane is parallel to the elongation of the OMC-1 
dust ridge at a 0.5 pc scale. The B$_{\bot}$ morphology at the 0.5 pc scale is found to be uniform \citep{vaillancourt08} 
and the axis of the OMC-1 ridge is perpendicular to the B field at that scale. 
% meaning?
Based on the symmetry properties, 
the B field at the larger 0.5 pc scale seems to guide the formation/condensation of the dense cores with their associated field structures at the smaller 20 mpc scale revealed with the SMA. 
Additionally, the stellar feedback from molecular outflows seems to further shape the B field.

The B field associated with the more evolved ultra-compact HII region G5.89-0.39 
shows more isolated patches (right panels in Figure 1 and \citet{tang09a}). The detected polarization is mostly 
toward the northern ridge. The dense structure revealed with smaller $\theta$ shows that 
there is an emission cavity in the southern ridge. This demonstrates that the polarization is sensitive to the 
underlying structure. The polarization angles vary significantly within 0.02 pc. 
Based on the analysis of the kinetic energy and the radiation pressure, 
we found that this region is active in stellar feedback. The B field is mostly overwhelmed. 

%\subsection{General properties}
The detected polarization emission often appears to be patchy. The depolarization zones are often 
associated with underlying structures or motions. 
We found that the highest polarization emission is not at the continuum peak. 
The connection of the B field to the larger scale parent cloud/envelope is not yet
conclusive due to the limited sample. For Orion BN/KL, a symmetry plane is identified and seems 
to connect to larger scale B field directions. For W51 e2, the influence of the larger scale B field 
is not clear, because the detected accretion plane and molecular outflows all deviate 
from the B field directions at larger scales. The B$_{\bot}$ morphologies are found to be 
more complex in the case where the source is more evolved (G5.89-0.39). This might suggest that the stellar 
feedback has a significant impact on the B$_{\bot}$ morphologies.
However, there are still limited sources observed with high angular resolution.

%Most of the polarization is parallel to the emission contours. We have further developed an analysis tool to extract the underlying physics. 

\section{Analysis Method: Magnetic Field Strength Maps}

The collapsing core W51 e2 \citep{tang09b} is the starting point for the 
polarization - intensity gradient method developed in \citet{koch12a}. In this 
new method, an ideal magneto-hydrodynamic (MHD) force equation is adopted and 
matched to observed dust Stokes $I$ and dust polarization maps. The observed 
systematic deviations between magnetic field and dust intensity gradient
orientations are further analyzed in that work. As demonstrated, the correlation
between the two orientations can be used to derive a magnetic field 
strength at each location where polarized emission is detected. The magnetic 
field strength $B$ as a function of position in a map can then be expressed as:
\begin{equation}
B=\sqrt{\frac{\sin\psi}{\sin\alpha}\left(\nabla P+\rho\nabla\phi\right)4\pi R},  \label{B}
\end{equation}
where $\rho$ and $P$ are the dust density and the hydrostatic pressure, and $\phi$ 
is the total gravitational potential. $R$ is the magnetic field radius. $\mathbf{\nabla}$ denotes the gradient.
The angle $\psi$ is the difference in orientations between the gravitational
pull and the intensity gradient, and the angle $\alpha$ is the difference between
the polarization and the intensity gradient orientations (Figure 3 in \citet{koch12a}).

\begin{figure}[h]
\includegraphics[scale=0.55]{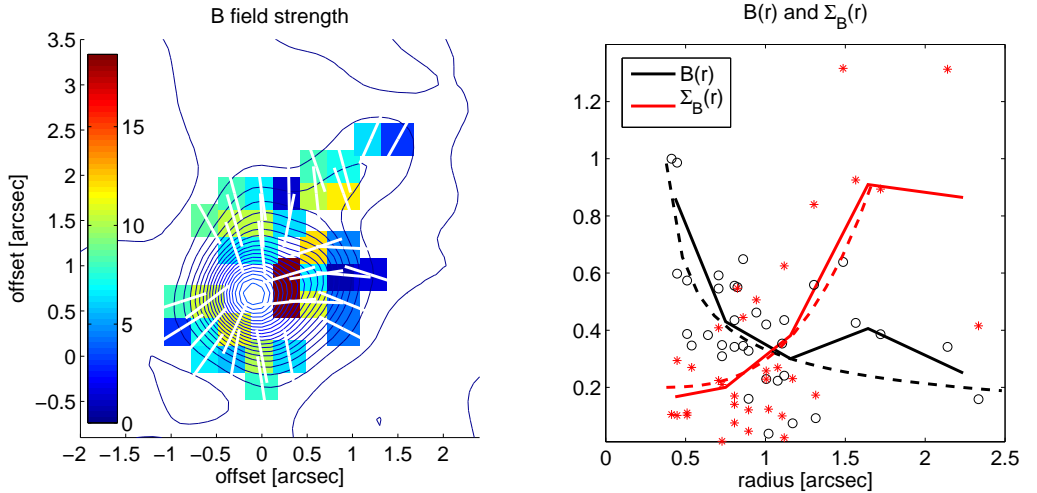}
\caption{\footnotesize Left Panel: Magnetic field strength map derived from the polarization - intensity
gradient method \citep{koch12a} for the collapsing core W51 e2. The units of the color coding are in mG. Overlaid in 
white are the magnetic field segments. 
The contours show the dust continuum Stokes I emission.
Right Panel: Magnetic field strength $B$ (black) and 
force ratio $\Sigma_B$ (red) as a function of radius from the emission peak. Open circles (o)
and asterisks ($\ast$) mark the individual data points. The solid lines are the azimuthally 
averaged values binned to half of the synthesized resolution. For comparison, the dashed lines
show scalings $\sim r^{-1/2}$ and $\sim r^{3}$ for the the profiles of the dimensionless field strength and the
ratio, respectively. The magnetic field strength is normalized to its maximum value of $\sim 18.7$~mG.
}
\label{field}
\end{figure}

As further demonstrated in \citet{koch12a}, the derived field strength is likely 
to be very little or not at all affected by projection effects. Thus, the method leads
to a total field strength. Reorganizing Equation (\ref{B}), the method also quantifies
the local magnetic field to gravitational force ratio, $\Sigma_B=\frac{\sin\psi}{\sin\alpha}$
(assuming $\nabla$P=0). It is remarkable that this
ratio only depends on two measured angles on an observed map. These angles represent the 
result of the imprint of the combined forces onto the dust and magnetic field morphologies. 
The ratio is free of any assumptions of mass and field strength or detailed modeling 
of the molecular cloud. 
The left panel in Figure \ref{field} shows the magnetic field strength map for W51 e2.
Displayed in the right panel are the azimuthally averaged force ratio $\Sigma_B(r)$ 
and the field strength profile $B(r)$. The field strengths vary between $\sim 2$~mG and  
$\sim 20$~mG with a radial profile $B(r) \sim r^{-1/2}$.

The local magnetic field to gravitational force ratio can also be converted to a 
beam-averaged differential mass-to-flux ratio \citep{koch12b}. A clear transition from 
a subcritical state at larger radii to a supercritical state closer to the center of 
the W51 e2 peak is found. Moreover, the force ratio $\Sigma_B$ can be interpreted in the 
context of diluted gravity \citep{shu97}. It, therefore, measures how efficiently 
the magnetic field is able to slow down the gravitational collapse. Based on this, 
a star formation efficiency reduced to $\sim 10$\% or less (compared to a free-fall
efficiency) is estimated in \citet{koch12b}.

We stress that -- although illustrated here for a molecular cloud core -- the polarization - intensity
gradient method is generally applicable to measurements that reveal the large-scale magnetic 
field morphology in combination with indications for gravity and/or pressure forces.  
The method can potentially also be expanded to Faraday rotation measurements in galaxies.

\section{Conclusion}

High-resolution dust polarization observations start to reveal detailed field structures in the star formation process. For further insight, it is paramount to develop analysis tools to quantitatively assess the role of the magnetic field.

%\subsection{Turbulence?}

%\acknowledgements{ The author is exceptionally grateful to anybody.}

\end{document}